
\magnification=\magstep1
\hsize=31pc
\vsize=43pc
\baselineskip=0.5cm
\def\p{k_{\parallel}}
\def\t{k_{\perp}}
\def\ni{\noindent}
\def\o{\omega}
\def\ms{\bigskip}
\def\1{\prime}
\def\2{\prime\prime}
\def\3{\prime\prime\prime}
\def\4{\prime\prime\prime\prime}

\def\ii{{\rm 1 \! l}}
\rightline{DFTT 6/1993}
\rightline{February 93}
\vskip 2.5truecm
\bf
\centerline{COLLECTIVE MODES IN A SLAB }
\medskip
\centerline{OF INTERACTING NUCLEAR MATTER:}
\medskip
\centerline{The effects of finite range interactions}
\ms\ms
\centerline{W.M. Alberico}
\rm
\centerline{Dipartimento di Fisica Teorica dell'Universit\`a--Torino,
 Italy}
\centerline{and INFN, Sezione di Torino, Italy}
\medskip
\bf
\centerline{P. Czerski}
\rm
\centerline{Institute of Nuclear Physics}
\centerline{ul. Radzikowskiego 152, Krak\'ow,Poland}
\medskip
\bf
\centerline{A. De Pace}
\rm
\centerline{Dipartimento di Fisica Teorica dell'Universit\`a--Torino,
  Italy}
\centerline{and INFN, Sezione di Torino, Italy}
\centerline{and}
\bf
\centerline{V.R. Manfredi}
\rm
\centerline{Dipartimento di Fisica dell'Universit\`a--Padova, Italy}
\centerline{and INFN, Sezione di Padova, Italy}
\ms
\vfill
\centerline{ABSTRACT}
\par\noindent
We consider a slab of nuclear matter and investigate the collective
excitations, which develop in the response function of the system.
We introduce a finite--range realistic interaction among the nucleons,
which reproduces the full G--matrix by a linear combination of gaussian
potentials in the various spin--isospin channels. We then analyze the
collective modes of the slab in the $S=T=1$ channel: for moderate
momenta hard and soft zero--sound modes are found, which exhaust most
of the excitation strength. At variance with the results obtained with
a zero range force, new ``massive'' excitations  are found for the
vector--isovector channel .

\vfill\eject

\ni{\bf 1. Introduction.}
\bigskip

The slab of interacting nuclear matter has revealed itself as an
interesting tool[1,2] for investigating  specific properties of nuclei:
being confined in one dimension only, it allows to consider, e.g., the
influence of a surface (already present in the semi--infinite slab of
Esbensen and Bertsch[3]) together with the effects of discrete levels,
which are appropriate for a spatially confined system. But at the same
time it keeps typical features of an infinite system, like for example
density oscillations with phonon--like dispersion relation, together
with translational invariance along the two unconfined directions.

This model also offers a schematic (and rather extreme) situation to be
compared with the one of highly deformed nuclei: as already pointed out
in ref.[2], the observed splitting of the dipole resonance can be
related with the analogous phenomenon which shows up in the collective
modes (zero sound) of the slab.

In this paper we consider, as in ref.[2], the response function of a
slab of interacting nucleons within the framework of the Random Phase
Approximation (RPA): however, while in the previous work a schematic
zero range force (of the Landau--Migdal type) was employed, here we
utilize a more realistic nucleon--nucleon interaction, which is derived
by suitably fitting previous G--matrix calculations [4,5]. The effective
interaction derived in the present work has the appropriate finite range
in all spin--isospin channels and allows a more detailed investigation
of the interplay between range of the interaction and size of the system.
This point will be explored in connection with the collective modes of
the slab, as derived within the RPA framework: indeed we will show how
the corresponding response function is modified in the presence of finite
range forces.

In Section 2 we briefly describe the model, summarizing the appropriate
formalism to deal with the response function of the slab; in Section 3
we give the derivation of the effective particle--hole interaction,
which is considered in the various spin--isospin channels. Finally in
Section 4 the resulting RPA response function and, specifically, the
collective modes of the slab are presented and shortly discussed.

\bigskip\bigskip\bigskip
\bf
\ni 2. Description of the model.
\bigskip
\rm
We consider, as in ref.[2], a system of noninteracting
fermions, confined in the domain $0 \leq z \leq L$ by an infinite
potential well; the single particle wave functions are then:

$$<\vec r|\vec k> = \sqrt{2\over SL} exp\left[i \vec \p
 \cdot \vec r_{\parallel} \right]
 \sin(\t z) \chi_{s} \xi_{t}~~,
 \eqno(2.1)$$

\ms
\ni where S is the area of the surfaces which set the boundaries of the
system (the slab), $\chi_{s}$ and $\xi_{t}$ are two-component
spinors in the spin and isospin space and the symbols $\parallel$
and $\perp$ correspond to the
 parallel and perpendicular directions to the slab surfaces.
 In (2.1), since at the end we shall let  $S \to \infty$,
 $\p$ is assumed to be a continuum wave number whereas $\t$ is
quantized according to

$$ \t = n {\pi \over L} = n {k_F \over M}~~,  \eqno(2.2) $$

\ms \ni with n=1,2,3,..., the total particle wave number being
 $k=\sqrt{ \p^2 +\t^2 }$. Eq.(2.2) also relates $M$ with the slab
thickness $L$ and to the Fermi momentum $k_F$: since $M$ has to be
an integer number, by fixing the slab thickness $k_F$ turns out to
be quantized (or viceversa, indeed for convenience we will assume a
fixed value for $k_F=1.36$~fm$^{-1}$).

{}From the single particle wavefunctions and (kinetic energy) eigenvalues
one can obtain the polarization propagator (or particle--hole Green
function) for the non--interacting Fermi gas of nucleons in the slab;
it reads

 $$\Pi^0(x_1,x_2) = - {4 i \over \hbar} G^0(x_1,x_2)
 G^0(x_2,x_1) \eqno(2.3)$$

\ms \ni  where $G^0$ is the free single particle Green's function and
the factor of 4 arises from the traces over spin and isospin (we assume
an isospin symmetric, N=Z, nuclear matter). The Fourier transform for
the slab system is then defined as follows:

$$\eqalignno{f(\vec \p,\t,\t^{\1}) &=
S \int d^2r_{\parallel}\, exp\left[ -i \vec \p \cdot
 \vec r_{\parallel} \right] &\cr
&\quad\times
\int\limits_0^L dz_1
\int\limits_0^L dz_2\, {\widetilde f}(\vec r_{\parallel},z_1,z_2)
 \cos(\t z_1) \cos(\t^{\1} z_2) &(2.4)\cr}$$

$$ \eqalignno{{\widetilde f}(\vec r_{\parallel},z_1,z_2) &=
{ 1 \over ( 2 \pi)^2 } {4\over L^2}  {1\over S}
 \int d^2\p \, exp\left[ i \vec \p \cdot r_{\parallel} \right]&\cr
&\quad\times
\sum_{\t,\t^{\1}}
f(\vec \p,\t,\t^{\1})\cos(\t z_1)\cos(\t^{\1} z_2)\eta_{\t}\eta_{\t^{\1}}
&(2.5)\cr}$$

\bigskip
 \ni where $ \vec \p $ should be thought of as $\vec \p - \vec \p^{\1}$
 since the system is translational invariant in the direction parallel
to the surface and $\eta_{\t} = 1 (1/2)$ for $\t \ne 0$ ($\t =0$).
By applying the above transformations to the polarization
propagator (for details and explicit formulas see ref.[2]) we calculate
the real and imaginary part of $\Pi^0_{slab} ( \vec \p,\t,\t^{\1},\o)$.

We remind that the imaginary part of the polarization
 propagator is directly related to the response function of the system,
thus providing the excitation strength of the slab when this is coupled
to an external probe; if the latter, for example,
induces charge density fluctuations into the system, the response per
particle is given by the relation:

$$ R( \p,\t,\o) =
- {2 \over N \pi}{\rm Im} \Pi^0_{slab} (\p,\t,\t,\o)~~ .
\eqno(2.6) $$

As already stated in the introduction,
in this work we are mainly concerned with the effects of a realistic
interaction on the slab response function, and with the interplay between
the finite range of the force and the finite size of the system. These
items are best investigated within the framework of the Random Phase
Approximation ( RPA ) scheme. We have already pointed out in ref.[2] the
main differences between the infinite nuclear matter and the slab RPA
equations for the polarization propagator. Due to the finite size of the
slab in the transverse dimension, translational invariance is lost and
the simple algebraic RPA equation[6] for the polarization propagator in
nuclear matter turns into a set of coupled equations for the
non--diagonal elements of $\Pi^{\rm RPA}_{\rm slab}$ in momentum space.

In order to derive the latter it is convenient to start with the RPA
equation for the polarization propagator in r--space; it reads (the
subscript ``slab'' is omitted to simplify the notation):

$$\eqalignno{
 \Pi^{\rm RPA}&(\vec r_1-\vec r_2,z_1,z_2)=
\Pi^0( \vec r_1 -\vec r_2,z_1,z_2 ) +
\int\limits\!d^2r_3\,d^2r_4 \int\limits_0^L\!dz_3\,dz_4 &\cr
&\quad\times
\Pi^0 ( \vec r_1 - \vec r_3,z_1,z_3 ) V(\vec r_3 - \vec r_4 ,z_3,z_4)
 \Pi^{\rm RPA}(\vec r_4 - \vec r_2,z_4,z_2 )&(2.7)\cr} $$

\ms\ni where the $\vec r_i$ are two dimensional vectors in the
(x,y)--plane.
Since the interaction is time independent, we shall not explicitly
indicate the time (or energy) dependence in our formulas.

Applying now the cosine Fourier transformation, defined in (2.4) and
(2.5), to both sides of (2.7) we get, after some manipulations, the
compact expression

$$\eqalignno{
 \sum_{\t^{\3}} &\left[ \delta_{\t,\t^{\3}} - {4 \over L} \rho_{slab}
\sum_{\t^{\2}} \eta_{\t^{\2}}\eta_{\t^{\3}}
\Pi^0(\p,\t,\t^{\2}) V(\p,\t^{\2},\t^{\3})\right]
\Pi^{RPA}(\p,\t^{\3},\t^{\1}) &\cr
&\qquad\qquad\qquad\qquad\qquad\qquad\qquad
 = \Pi^0(\p,\t,\t^{\1}) ~~. &(2.8)\cr} $$

This equation is easy to solve numerically since the transverse
momenta are discretized according to $\t=nk_F/M$ ($n=0,1,2\dots$)
and the longitudinal momentum $\p$ can take arbitrary values and plays
the role of a fixed parameter. For a definite choice of $\p$ and $\o$
(which is implicit in the free and RPA polarization propagators),
eq. 2.8 becomes a two--dimensional system of linear algebraic equations.
We cut the range of the allowed $\t$-values when it exceeds by more than
three times the upper limit of the corresponding response region for the
free system. Moving to $\t$ values higher that $\t^{max}$ does not affect
the results by more than one percent and doesn't change the qualitative
features of the solution.

Concerning the particle--hole interaction $V(\p,\t^{\1},\t^{\2})$
between nucleons in the slab system, it is derived from a finite range
realistic NN potential, as it will be explained in the next Section.
We anticipate here that in nuclear matter it is represented by a
linear combination of gaussian functions both in momentum space,
$$V(k)=g\,e^{-k^2/m^2}, \eqno(2.9)$$
where $g$ and $m$ are fitting parameters, and in r--space, where it
reads, correspondingly,
$$V(r)=Ce^{-\mu r^2}, \eqno(2.10)$$
with $\mu=m^2/4$, $C=g (m^3/8 \pi^2) \sqrt{\pi}$.

For the slab system we have to perform the Fourier cosine
transformation (2.4) on gaussian functions like (2.10) and the resulting
particle--hole force (apart from the spin--isospin matrix elements)
turns out to be of the following form:
$$V(\p,\t^{\1},\t^{\2})= S C {\pi \over \mu} e^{\p^2 /
4 \mu} \int\limits_0^L \int\limits_0^L dz_1 dz_2 e^{-\mu (z_1-z_2)^2}
\cos(\t^{\1} z_1) \cos(\t^{\2} z_2),
\eqno(2.11)$$
where the integration over $z_1,z_2$ has to be carried out numerically.
As a result of the finite range of the interaction
$V(\p,\t^{\1},\t^{\2})$ is neither
diagonal with respect to $\t^{\1}$ and $\t^{\2}$ nor constant in
momentum space (at variance with ref.[2]).

\ms\ms
{\bf \ni
3. The Effective Interaction.}
\ms\ms

To proceed with the calculation of the RPA response function, we need
the finite range potential which acts between nucleons.
We start from any realistic NN potential derived by fitting the
NN scattering data and, in order to properly deal with the strong
short range correlations, we insert this ``bare'' interaction in the
Bethe--Goldstone equation. In the operator form the latter reads:
$$\rm G=V+V{Q \over E-H_0} G \eqno(3.1)$$
where V is the bare NN potential, G is the so called Brueckner G--matrix
(which will coincide with our effective interaction), E is the starting
energy, $\rm H_0$ is the Hamiltonian operator for the intermediate
two--particle states; finally Q is the Pauli operator, which takes care
of the medium effects of the Fermi system, by forbidding the particles
to scatter into occupied intermediate states.

The Bethe--Goldstone equation is then solved in the infinite nuclear
matter, following the method suggested by Haftel and Tabakin[7].
Eq.(3.1) implicitly depends upon the density of the system via the
Pauli operator; we fix the starting energy to be
E=74~MeV. Further details can be found in references [4,5]. As a
solution we have the so called G--matrix, which can then be expressed
in terms of the direct and exchange matrix elements of an effective
potential: the latter is subsequently represented by means of a
suitable parameterization.

The method we employ has been developed in ref.[5]: working in the
particle--hole representation, one interprets the resulting G--matrix
as direct plus exchange matrix elements of some effective potential,
whose dependence upon spin and isospin operators is explicitly taken into
account. Then the momentum dependence of this potential is suitably
parameterized in terms of simple Yukawa's functions, with the aim of
obtaining an effective potential which is vaguely related to the
exchange of some ``effective'' mesons.

For the purpose of applications to the slab system the Yukawa's form
of the interaction hinders the (partially) analytical evaluation of
the particle--hole force (2.11). Thus in the present work we have
considered a local parameterization of the G--matrix elements in the form
of simple gaussians. These might turn quite advantageous also for
finite nuclei calculations, e.g. when the single particle wave
functions are expressed in terms of Harmonic Oscillator eigenstates.

To illustrate the fitting procedure let us consider the particle--hole
matrix elements of G in a definite spin--isospin channel,
$<ST|G({\vec k},{\vec p_1},{\vec p_2})|ST>\equiv <ST|U_{ph}|ST>$; as
shown in Fig.1, the direct matrix element depends on the total p--h
momentum ${\vec k}$ while the exchange one depends upon
${\vec p_1}-{\vec p_2}$ (the center of mass dependence being taken into
account by some averaging procedure).

We now assume that the momentum dependence of the G--matrix,
solution of eq.(3.1) with some realistic input for the NN bare
potential, can be accurately
reproduced by a two--body potential of the form:
$$\eqalignno{U_{ph}({\vec q})&=
g_{00}\,e^{-q^2/m^2_{00}} \ii_{\sigma} ~~ \ii_{\tau} +
g_{01}\,e^{-q^2/m^2_{01}}\ii_{\sigma}({\vec\tau_1}\cdot{\vec\tau_2}) &\cr
& +
g_{10}\,e^{-q^2/m^2_{10}}({\vec\sigma_1}\cdot{\vec\sigma_2})\ii_{\tau}
+g_{11}\,e^{-q^2/m^2_{11}}({\vec\sigma_1}\cdot{\vec\sigma_2})
({\vec\tau_1}\cdot{\vec\tau_2}), &(3.2)\cr}$$
where for sake of simplicity we do not include tensor components,
which could be extracted as well by considering definite spin
projections in the G--matrix elements. Considering, for example, the
$S=0, T=0$ channel, the p--h matrix
element of (3.2) reads (with the same notation of Fig.1):
$$\eqalignno{&<00|U_{ph}|00>=g_{00}\left(4e^{-k^2/m^2_{00}}-
e^{-({\vec p_1}-{\vec p_2})^2/m^2_{00}}\right) &\cr
&\quad -3g_{01}\,e^{-({\vec p_1}-{\vec p_2})^2/m^2_{01}}
-3g_{10}\,e^{-({\vec p_1}-{\vec p_2})^2/m^2_{10}}
-9g_{11}\,e^{-({\vec p_1}-{\vec p_2})^2/m^2_{11}}. &(3.3)\cr}$$
With the approximation ${\vec p_1}-{\vec p_2}\simeq 0$
(which turns out to be accurate,
at least for not too large particle momenta) the
exchange contributions simply reduce to a constant and, in order
to fix the parameters in the effective potential $U_{ph}$, eq.(3.2),
one has to solve the following system:
$$\eqalign{ <00|G(k)|00> &=
g_{00}\left(4e^{-k^2/m^2_{00}}-1\right) -3g_{01}-3g_{10}-9g_{11}\cr
<01|G(k)|01> &=
-g_{00}+g_{01}\left(4e^{-k^2/m^2_{01}}+1\right)-3g_{10}+3g_{11}\cr
<10|G(k)|10> &=
-g_{00}-3g_{01}+g_{10}\left(4e^{-k^2/m^2_{10}}+1\right)+3g_{11}\cr
<11|G(k)|11> &=
-g_{00}-g_{01}+g_{10}+g_{11}\left(4e^{-k^2/m^2_{11}}-1\right)\cr}
\eqno(3.4)$$
This allows to extract strengths ($g_{\scriptstyle ST}$) and masses
($m_{\scriptstyle ST}$)
for all components in $U_{ph}$. In order to obtain a good fit to the
original solution two gaussians are needed for each spin--isospin
channel. Moreover we found it more convenient (and accurate) to find
separate parameterization for the full G-matrix elements and for
the direct part of them. For example, in the $S=T=0$ channel one sets:
$$<00|G|00>_{\rm tot}=4\left(g_{00,F}^{(1)}\,
\exp\{-k^2/({m^{(1)}_{00,F}})^2\}
+g_{00,F}^{(2)}\,\exp\{-k^2/({m^{(2)}_{00,F}})^2\}\right),
\eqno(3.5a)$$
$$<00|G|00>_{\rm dir}=4\left(g_{00,D}^{(1)}\,
\exp\{-k^2/({m^{(1)}_{00,D}})^2\}
g_{00,D}^{(2)}\,\exp\{-k^2/({m^{(2)}_{00,D}})^2\}\right).
\eqno(3.5b)$$

The parameters of the present fit are summarized in Table I
for the direct part of the G--matrix and in Table II for the full
interaction. We observe that, at variance with ref.[5], the
present parameterization involves the full G--matrix and not only
the corrections induced by the ladder diagrams on the original
bare potential (which in some cases are artificially large);
thus one does not need the explicit knowledge of the NN
interaction employed in the Bethe--Goldstone equation. The
effective particle--hole potential resulting from our fit is
illustrated in Fig.2, where the full p--h interaction is displayed
for each spin--isospin channel and compared with the exact G--matrix.
The quality of the fit is fairly good, but for the scalar isoscalar
channel, which however will not be utilized in the following: indeed
in this case the interaction turns out to be strongly attractive over
the whole range of momenta and cannot produce the hard collective modes
we are interested in.

\ms\ms
{\bf \noindent 4. Results and discussion.}
\ms\ms

We have solved the RPA equations (2.8) for the slab polarization
propagator by utilizing the effective potential of the previous
Section. As one can see from Fig.2 the full interaction develops a
large repulsive p--h force in the spin--isospin channel ($S=T=1$),
together with a rather strong momentum dependence; the latter (although
we are neglecting the tensor components of the interaction) should be
ascribed to the presence, in this channel, of the long range one pion
exchange potential. As it is well known[8] the attractive pion exchange
modulates the strong repulsive short range correlations in the
spin--isospin channel, thus producing some softening of the RPA
response function.

{}From the resulting momentum dependence of the p--h force
we expect indeed to observe a different behaviour, in the
collective response of the system, with respect to ref.[2], where
a zero--range (constant in momentum space) interaction was used
within the same theoretical framework. One should also remind that the
interaction utilized here is not diagonal in the perpendicular momentum,
thus involving a larger influence of the non--diagonal terms of $\Pi$
in the RPA equations (2.8).

Let us then focus on the vector--isovector channel: it is well
known that, in nuclear matter, at small energy and momentum
transfers the repulsive
interaction is strong enough to produce a collective excitation
(zero sound) outside the continuum p--h response, which in turn is
severely depressed.

In the slab system and with the present finite range interaction
we have found, in analogy with ref.[2], two distinct collective
modes, according whether the transfer momentum is parallel to the
infinite dimensions of the system or perpendicular to it: in
Fig.3a,b we display these zero sound dispersion relations, which
have approximately a linear form: $\o=v k$, $v$ being the
corresponding velocity of propagation of the bosonic excitation
in the nuclear medium. Results are displayed for infinite nuclear
matter and for slabs with two different thicknesses (which are
fixed by the value of the integer $M$). Fig.3b shows a magnified view
of Fig.3a at somewhat larger momentum transfers, where one can better
distinguish the different modes occurring for the $M=2$ and $M=8$
slabs.

More precisely, the sound
velocity in nuclear matter turns out to be $v_s^{n.m.}=1.62 v_F$ (it
would be $v_s^{n.m.}=1.71 v_F$ with a $\delta$--force of approximately
the same strength as the one of the gaussian interaction at $q=0$);
$v_F=0.286 c$ is the Fermi velocity.
For the $M=2$ slab the two sound velocities
obtained with our gaussian interaction are (in parenthesis the
corresponding values for the $\delta$--force) $v_{\parallel}=1.36 v_F$
($1.5 v_F$) and $v_{\perp}=1.54 v_F$ ($1.89 v_F$), for the parallel
and perpendicular modes, respectively. In a thicker slab ($M=8$) the
resulting sound velocities are closer to each other and to the infinite
nuclear matter value: $v_{\parallel}=1.56 v_F$ ($1.67 v_F$) and
$v_{\perp}=1.58 v_F$ ($1.79 v_F$).

With respect to the analogous outcome in ref.[2] one should
notice the following qualitative differences :

\item{i)} the dispersion
relation in nuclear matter is modified by the present
interaction, since as the momentum increases the p--h force
weakens (see Fig.2) thus producing some softening with respect to
a constant interaction;
\item{ii)} for a fixed thickness of the slab,
one still finds two distinct collective modes, the faster of
which being associated with a transverse momentum transfer
(``perpendicular'' mode), however {\it both} the perpendicular
and the parallel modes lie {\it below} the nuclear matter curve,
while in the previous work they were sitting on opposite sides.
This outcome should signal the role played by the off--diagonal
(in the perpendicular momentum) terms of the interaction, which were
absent for the $\delta$--force and produce a sizeable softening
both of the longitudinal and, even more, of the transverse modes.
\item{iii)} we remark that the above discussed
softening is due to the attractive part of the interaction and
this, in turn, has to be associated with pion exchange, namely with
the longest range component of the NN force. It is thus worth
noticing that the relative lowering of the transverse zero sound
velocity is much larger than the one of the parallel mode: this fact
signals an important correlation between the range of the interaction
and the transverse dimension of the slab, where
the finite size of the system plays the major role.

Beside the previously illustrated collective modes, the present model
displays additional excitations, which in the RPA scheme are signalled
by poles in ${\rm Re}\Pi^{RPA}$, not occurring in $\Pi^0$,
corresponding to very narrow peaks in ${\rm Im}\Pi^{RPA}$ (and thus in
the response function); this is shown in Fig.4, for the $M=8$ slab.
The energy--momentum dispersion relation of these
modes, which occur at higher energies with respect to the zero sound,
differs from the latter since it does not vanish in the zero momentum
limit and then stays almost constant, with a weak quadratic dependence.
They are illustrated in Fig.5, for the $M=2$ and $M=8$ slabs: in the
thicker system the energy interval between any two of these modes is
smaller (they appear to be almost equally spaced); however the strength
in the corresponding peaks (see Fig.4) rapidly goes down. Indeed they
disappear in the infinite system.

The origin of these collective excitations lies in the finite range of
the interaction utilized here and critically depends upon the interplay
between range of the interaction and extension of the system: they did
not show up when a zero--range force was employed, as in ref.[2].
Moreover the peaks seem to be more pronounced when the range of the
interaction (2-3 fm) is comparable with the transverse dimension of the
slab: in the $M=8$ slab ($L\simeq 18$~fm) only the lowest energy mode
displays a significant strength.

This situation resembles the low frequency modes occurring in crystals
and in spin systems (ferromagnets, antiferromagnets, etc.): in both
cases one observes (one or more) acoustical and optical branches, the
former having vanishing frequency when $k\to 0$, the latter requiring
finite excitation energy in the long wavelength limit. In a ferromagnetic
system the above mentioned excitations are spin waves usually called
magnons[9]: experimental evidence for these modes has been revealed
by inelastic neutron scattering[10]. From the microscopic point of view,
they could be explained in terms of collective excitations of an
infinite range Ising model[11], which allows to associate the finite
frequency of the optical branch to the order parameter of the
ferromagnetic phase.

In the present case we notice that the above mentioned ``massive'' modes
have been found in the spin--isospin channel ($S=T=1$) and may be
explored by letting the system to interact with an external
electromagnetic field.
They arise only when the interaction has a {\it non--zero range} and
disappear when translational invariance is restored, by letting
$L\to\infty$;
indeed in infinite nuclear matter the zero sound alone is allowed as a
collective mode, irrespective on the (finite or zero) range of the model
interaction utilized. It is worth reminding that the only physical
example of a similar mode occurring in an infinite system, the electron
gas, are the so--called plasmons, whose origin is driven by the infinite
range of the Coulomb potential. Here instead we have finite range forces
and the ``plasmon--like'' excitations survive only as far as the range
of the interaction is comparable with the slab thickness $L$.
Actually in the infinite system the three
different types of  collective excitations which we have found
(parallel and perpendicular zero sound, and plasmon--like) collapse into
a single collective mode, with linear dispersion relation.

It might be interesting to notice that in a Wigner lattice the occurrence
of a similar situation has been interpreted in connection with the
phenomenon of spontaneous symmetry breaking. This model displays three
different types of elementary excitations:
two transverse modes which behave as phonons, with $\omega\propto q$,
and a longitudinal collective mode (the ``plasmon'') which has a finite
frequency as $q\to 0$[12--14]. According to Anderson the plasmon can be
viewed as the massive ``Higgs'' boson associated with the spontaneous
breaking of gauge invariance which is due, in the presence of the
Coulomb potential, to the density fluctuations in the electron gas.
Notably, the existence of this massive mode is crucially related to the
long range of the Coulomb interaction. Short range correlations, like
in crystals, would give rise to massless Goldstone bosons alone.

As a final remark we would like to point out that the present
calculation, although rather crude in the description of the nuclear
confinement, yet utilizes a realistic p--h force, thus allowing to bear
some confidence in the qualitative (if not quantitative) results we
obtain for the collective excitations of the system.
In particular we believe that
the analysis of the interplay between the range of the interaction and
the finite size of the system might offer some hints also for
experimental research in highly deformed nuclei.

\bigskip\bigskip\bigskip
Acknowledgements\hfill\break
The authors wish to thank Prof. A. Molinari for fruitful discussions.

\bigskip
\bigskip
Work was partially supported by Grant No. KBN 202049101.

\bigskip\bigskip\bigskip
\bigskip\bigskip\bigskip
\vfill\eject

\ni REFERENCES.

\bigskip
\item{[1]} W. M. Alberico, A. Molinari and V. R. Manfredi, Phys.Lett.
{\bf B194}, 1-5 (1987);
\item{[2]} W.M. Alberico, P. Czerski, V.R. Manfredi and A. Molinari,
Zeit. f\"ur Physik {\bf A338}, 149-155 (1991);
\item{[3]} H. Esbensen and G. F. Bertsch, Annals of Physics {\bf 157},
255-281 (1984);
\item{[4]} W.H.~Dickhoff, Nucl. Phys. {\bf A399} (1983) 287;
\item{[5]} P.~Czerski, W.H.~Dickhoff, A.~Faessler, and H.~M\"uther,
Nucl. Phys. {\bf A427} (1984) 224;
\item{[6]} see, for example, A. L. Fetter and J. D. Walecka,
 {\sl Quantum Theory of Many-Particle systems} (McGraw-Hill, New York,
1971);
\item{[7]} M.I. Haftel and F. Tabakin, Nucl. Phys. {\bf A158} (1970), 1;
\item{[8]} W.M.Alberico, M.Ericson and A.Molinari, Phys. Lett. {\bf 92B}
(1980), 153;
\item{[9]} C.Kittel, {\it Quantum theory of solids}, John Wiley and Sons
(1987);
\item{[10]} Brockhouse and Watanabe, IAEA Symposium,
Chalk River, Ontario, 1962;
\item{[11]} J.W. Negele and H.Orland, {\it Quantum Many--Particle
Systems}, Addison--Wesley Pub. Co. (1988);
\item{[12]} P.W.Anderson, Phys. Rev. {\bf 130} (1963), 439
\item{[13]} P.W.Anderson, {\it Basic Notions of Condensed Matter Physics}
Frontiers in Physics, (1984) Benjamin/Cummings Publ. Co.
\item{[14]} E.Wigner and F.Seitz, Phys. Rev. {\bf 43} (1933), 804;
{\bf 46} (1934), 509.
\vfill\eject
\nopagenumbers
\advance\vsize 26.1truept
\midinsert

\medskip

\centerline {\bf Table I}

$$ \vbox{\offinterlineskip \tabskip=0pt
\halign
{ \strut \vrule \quad#
& \quad \hfil # \hfil \quad &  \vrule#&  \quad \hfil #   \quad
&  \vrule# &  \quad \quad \hfil # \quad & \vrule#
&  \quad \hfil # \quad & \vrule# &  \quad \hfil # \quad &
 \vrule #
\cr
\noalign{\hrule}
& \multispan9 \hfil Direct Matrix Element \hfil & \cr
\noalign{\hrule}
& && && && && & \cr
&  && $g^{(1)}_{ST}$ && $m^{(1)}_{ST}$ && $g^{(2)}_{ST}$ &&
  $m^{(2)}_{ST}$ & \cr
& S T && [$MeV fm^3$] && [$MeV$] && [$MeV fm^3$] && [$MeV$] & \cr
& && && && && & \cr
\noalign{\hrule}
& && && && && & \cr
& 0 0 && -1014.40 && 4379.0 && 423.25 && 313.6  & \cr
& && && && && & \cr
& 0 1 && 187.45  && 449.7  && 425.23 && 2773.5 & \cr
& && && && && & \cr
& 1 0 && -426.00  && 4478.2 && 163.00  && 654.8  & \cr
& && && && && & \cr
& 1 1 && -266.24 && 9553.0 && 272.70  && 370.8  & \cr
& && && && && & \cr
\noalign{\hrule}}}$$
\rm
\smallskip
\noindent Table I -- Parameterization of the Direct p--h matrix elements
of the G--matrix of ref.[4,5]; $S,T$ denote the total spin and isospin
quantum numbers of the relevant channels.

\bigskip

\centerline {\bf Table II}

$$ \vbox{\offinterlineskip \tabskip=0pt
\halign
{ \strut \vrule \quad#
& \quad \hfil # \hfil \quad &  \vrule#&  \quad \hfil #   \quad
&  \vrule# &  \quad \quad \hfil # \quad & \vrule#
&  \quad \hfil # \quad & \vrule# &  \quad \hfil # \quad &
 \vrule #
\cr
\noalign{\hrule}
& \multispan9
\hfil Direct+Exchange Matrix Element \hfil & \cr
\noalign{\hrule}
& && && && && & \cr
&  && $g^{(1)}_{ST}$ && $m^{(1)}_{ST}$ && $g^{(2)}_{ST}$ &&
$m^{(2)}_{ST}$ & \cr
& S T && [$MeV fm^3$] && [$MeV$] && [$MeV fm^3$] && [$MeV$] & \cr
& && && && && & \cr
\noalign{\hrule}
& && && && && & \cr
& 0 0 && -2865.64 && 862.0 && 185.90 && 862.0  & \cr
& && && && && & \cr
& 0 1 && 576.20  && 10000.0  && 576.2 && 377.0 & \cr
& && && && && & \cr
& 1 0 && 321.14  && 9691.0 && 606.16  && 669.8  & \cr
& && && && && & \cr
& 1 1 && 1307.6 && 461.0 && -388.74  && 1363.0  & \cr
& && && && && & \cr
\noalign{\hrule}}}$$
\rm
\smallskip
\noindent Table II -- Parameterization of the Direct+Exchange p--h
matrix elements of the G--matrix of ref.[4,5];
notations are the same as in Table I.

\endinsert

\vfill\eject
\vskip 1.5 truecm
\centerline{\bf Figure Captions}
\bigskip
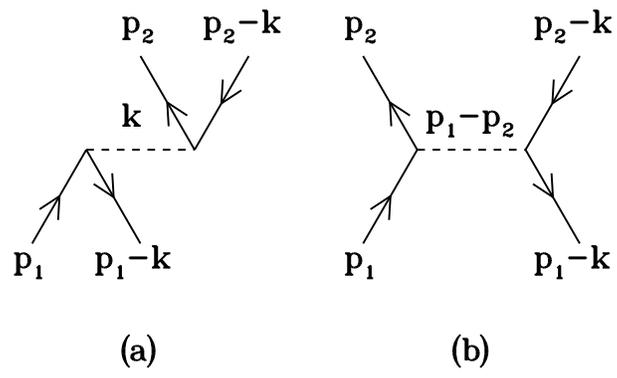
\item{}Fig.~1 -- Graphical representation of the direct (a) and
exchange (b) particle--hole matrix element of the interaction.
\smallskip
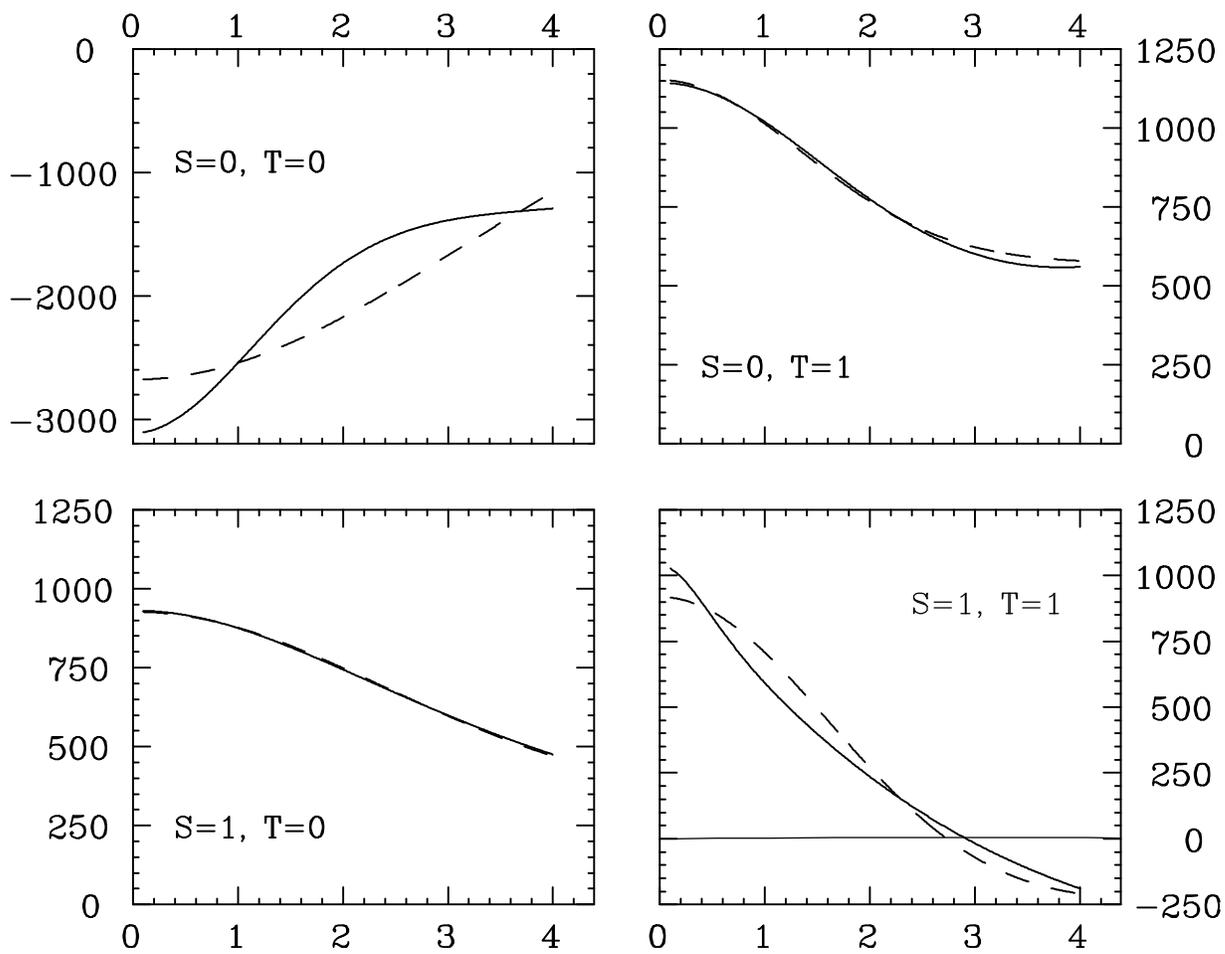
\item{}Fig.~2 -- The present gaussian fit (continuous lines) to the
p--h matrix elements of the full (direct plus exchange) G--matrix
(dashed lines) is presented as a function of $k$ (in fm$^{-1}$) for
the various spin--isospin channels. The units on the vertical scales
are MeV~fm$^3$.
\smallskip
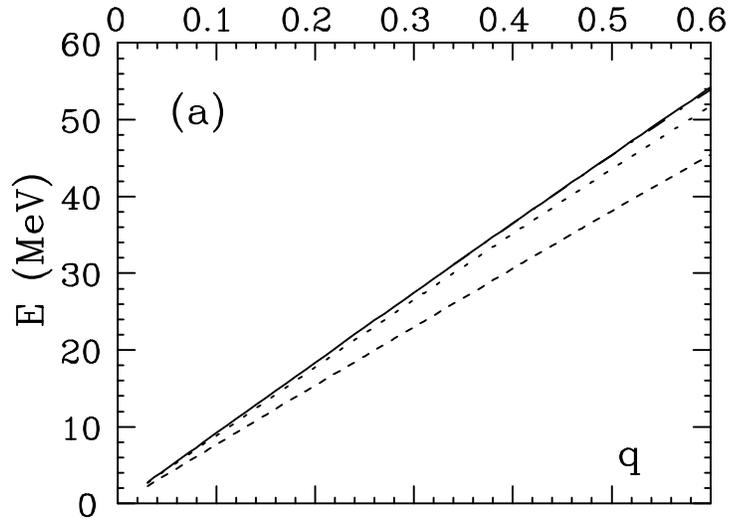
\item{}Fig.~3 -- The zero sound dispersion relations for the infinite
nuclear matter (continuous line) and for two slabs with $M=2$ and $M=8$:
the parallel modes correspond to the short--dashed ($M=2$) and dotted
($M=8$) lines, while the transverse modes (which are not
visible in upper part (a) of the figure) are represented by the dashed
and dot--dashed curves, respectively. The latter must be considered as
interpolations, to guide the eye, of the discrete--$q$ points where the
transverse zero sound can be found.
\smallskip
\item{}Fig.~4 -- Re$\Pi^{\rm RPA}$ (dotted line) and Im$\Pi^{\rm RPA}$
(continuous line) for the $M=8$ slab, at $q_{\parallel}=0.4$~fm$^{-1}$,
$q_{\perp}=0$, versus energy.
\smallskip
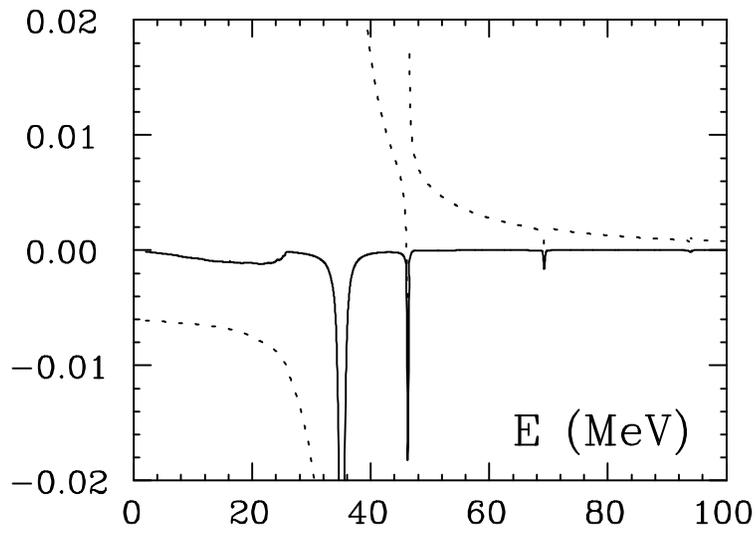
\item{}Fig.~5 -- Energy versus momentum behaviour of the collective
modes in the parallel direction for the $M=2$ slab (dashed lines) and
the $M=8$ slab (dotted lines); the infinite nuclear matter zero sound
is also displayed (continuous line).

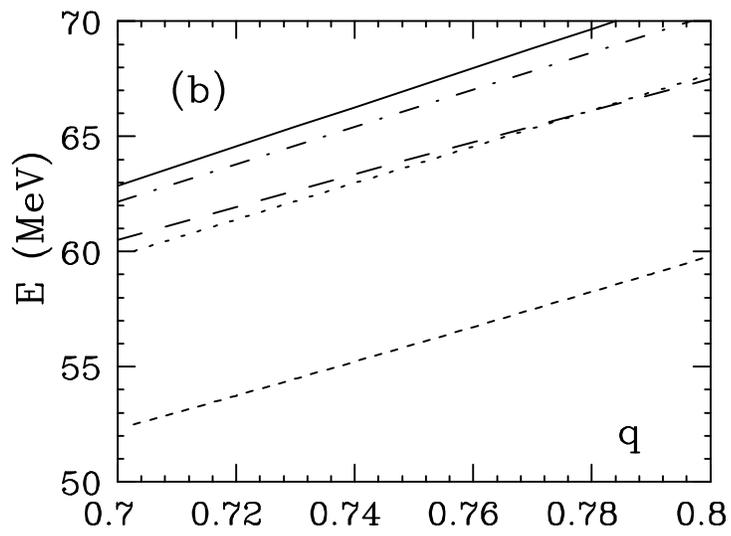
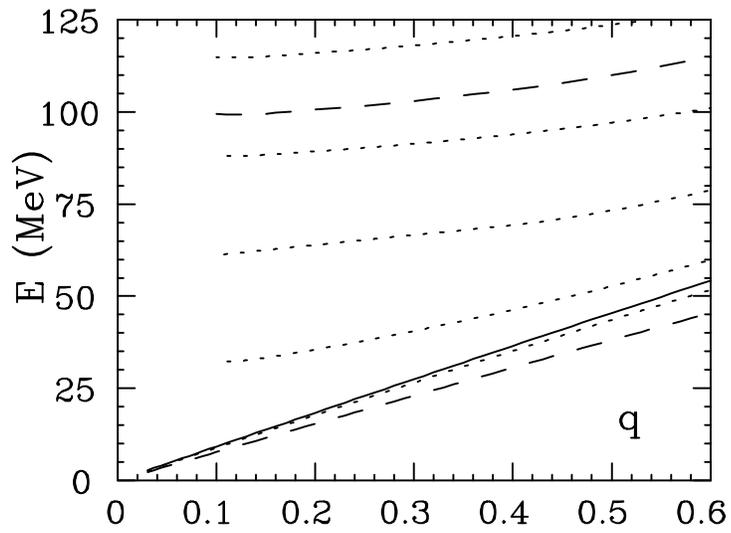
\bye